\documentclass[12pt]{article}

\usepackage{amssymb,amsmath,mathrsfs,epstopdf,slashed,color}
\usepackage{cite}
\usepackage{hyperref}

\allowdisplaybreaks[0]

\makeatletter

\@addtoreset{equation}{section}
\makeatother

\DeclareMathOperator*{\tr}{tr}

\title{\bf A Note on No-go in CY}
\author{Gary Shiu and Yoske Sumitomo}

\begin{document}

\begin{titlepage}

\setcounter{page}{0}
  
\begin{flushright}
\small
 MAD-TH-11-06\\
 TIFR/TH/11-31
\date \\
\normalsize
\end{flushright}

\vskip 3cm
\begin{center}
    
  {\Large \bf Stability Constraints on Classical de Sitter Vacua}
  
\vskip 2cm
  
{\large Gary Shiu${}^{1,2}$ and Yoske Sumitomo${}^3$}
 
 \vskip 0.7cm

 ${}^1$ Department of Physics, University of Wisconsin, Madison, WI 53706, USA\\
 ${}^2$ Institute for Advanced Study, Hong Kong University of Science and Technology, Hong Kong\\
 ${}^3$ Tata Institute of Fundamental Research, Mumbai 400005, India

 \vskip 0.4cm

Email: \href{mailto:shiu@physics.wisc.edu, yoskesumitomo@gmail.com}{shiu at physics.wisc.edu, yoskesumitomo at gmail.com}

\vskip 1.5cm
  
\abstract{\normalsize
 We present further no-go theorems for
 classical de Sitter vacua in Type II string theory, i.e., de Sitter constructions that do not invoke 
 non-perturbative effects or explicit supersymmetry breaking localized sources.
 By analyzing the stability of the 4D potential 
 arising from compactification on manfiolds with curvature, fluxes, and orientifold planes, we found that additional ingredients, beyond the minimal ones presented so far, are necessary to avoid the presence of unstable modes.
 We enumerate the minimal setups for (meta)stable de Sitter vacua to arise in this context.
  }
  
\vspace{4cm}
\begin{flushleft}
 \today
\end{flushleft}
 
\end{center}
\end{titlepage}

\setcounter{page}{1}
\setcounter{footnote}{0}

\tableofcontents

\parskip=5pt

 \section{Introduction}

The observational evidence for an accelerating universe adds another serious wrinkle to
the already vexing issue of moduli stabilization in string theory.
Besides having to stabilize a myriad of moduli self-consistently at values that give physically acceptable
couplings, 
requiring the vacuum solution to have 
positive
energy introduces a new layer of complications.
Unlike their anti de Sitter and Minkowskian counterparts, de Sitter solutions
are much harder to construct as they
are not amenable to the powerful tools of supersymmetry.
Furthermore, the de Sitter solutions one might hope to obtain from a string compactification
 are at best metastable. There  exist  generically
 supersymmetric vacua in the
 decompactified limit and any candidate de Sitter vacua are subject to all kinds of perturbative and non-perturbative instabilities.

In the past decade, motivated partly by the increasing 
observational support for dark energy, various proposals for constructing metastable de Sitter vacua from string theory have been suggested. These proposals can be broadly divided into two types: those that hinges on non-perturbative effects and/or explicit SUSY breaking localized sources (most notably \cite{Kachru:2003aw} and variations thereof), and those that do not \cite{Hertzberg:2007wc,Silverstein:2007ac,Haque:2008jz,Danielsson:2009ff,Danielsson:2011au,Caviezel:2008tf,Flauger:2008ad,deCarlos:2009fq,deCarlos:2009qm,Caviezel:2009tu,Dong:2010pm,Andriot:2010ju,Dibitetto:2010rg,Danielsson:2010bc,Wrase:2010ew}.
 We refer to the latter as {\it classical de Sitter solutions}, as their constructions
 involve only classical ingredients such as internal curvature, fluxes, and orientifold planes.
 In contrast to non-perturbative effects in string compactifications
  which are
  difficult to compute in full detail, 
the simplicity of these classical de Sitter 
solutions
allows for 
{\it explicit} models to be constructed.
Among these models, some of them can be shown to be genuine solutions of the 10D equations of motion, as well as critical points of the dimensionally reduced theory.
The classical de Sitter solutions explored so far are admittedly far from realistic,
e.g., the Standard Model sector has not yet been implemented and 
an exponentially small cosmological constant is difficult to achieve. 
Nonetheless,  they serve to illustrate the general issues one may 
encounter in 
constructing explicit de Sitter vacua from string theory.
Furthermore, 
explicit constructions of such solutions, though not fully realistic, may 
shed light on conceptual issues of de Sitter space, such as a microscopic understanding of its entropy and holography etc.

If one may draw hints from the aforementioned attempts in constructing explicit de Sitter vacua,
a recurrent lesson seems to be the ubiquity of tachyons.
While anti de Sitter  flux vacua are abundant \cite{DeWolfe:2005uu}
\footnote{The non-perturbative (in)stability of $AdS$ vacua were discussed in \cite{Narayan:2010em,Harlow:2010az}.
See also the non-perturbative instability argument \cite{Horowitz:2007pr} for $AdS_5$ through a bubble of nothing.}
, de Sitter solutions are hard to come by. Even if de Sitter extrema are found, they are plagued with one or more unstable modes.
Similar searches for de Sitter vacua within 4D supergravity also seems to suggest that tachyonic modes are omnipresent \cite{Covi:2008ea,GomezReino:2008bi,Borghese:2010ei,Brizi:2011jj}, though the tachyons found in explicit reductions of 10D backgrounds (e.g., \cite{
Danielsson:2011au,Danielsson:2010bc})
are not necessarily in 
the ``sGoldstino'' direction \cite{Wrase:2010ew}.
Taking cues from these earlier attempts, we set out to prove some no-go theorems for the existence of stable de Sitter vacua.
Of course, no-go theorems always come with assumptions and there are ways around them. Our
work thus helps in sharpening the requirements needed for constructing explicit de Sitter vacua from string theory.

We found that some necessary constraints on the absence of tachyonic modes in classical de Sitter solutions
can be stated in a 
surprisingly clean and simple way.
From the scalings of various contributions to the potential with respect to the universal moduli (i.e., dilaton and the breathing mode), and upon simplifications around a positive potential extremum, we analyzed the conditions under which the 
moduli mass matrix
 contains necessarily an unstable mode or a flat direction upon diagonalization.
The conditions we found, supplemented with earlier no-go theorems on the existence of de Sitter extrema \cite{Hertzberg:2007wc,Wrase:2010ew,Haque:2008jz,Danielsson:2009ff}, therefore provide us with a more refined guide to search for (meta)stable de Sitter vacua in classical supergravity.
As we shall see, the conditions on the stability
of de Sitter extrema are simple but yet powerful enough
to show that the minimal setups evading the no-goes in
\cite{Hertzberg:2007wc,Haque:2008jz,Danielsson:2009ff,Wrase:2010ew} turn out to all suffer from perturbative instabilities. 
To obtain stable de Sitter vacua, additional ingredients (e.g., more types of fluxes and/or O-planes) have to be introduced.
Our approach further allows us to enumerate the {\it minimal} ingredients needed for constructing classical de Sitter vacua.

Although our focus is on finding classical vacua with positive cosmological constant in this paper, we should mention here some recent related attempts in clarifying the possibilities of realizing a
time-varying dark energy in string theory.
A shift-symmetry is often invoked to prevent any undesired couplings between the heavy modes and the quintessence field.
There are two strands of approaches to weakly break this shift symmetry in order to obtain a
 time-varying dark energy: one with a non-perturbative potential \cite{Choi:1999xn,Svrcek:2006hf}, and the other with a classical potential from an NS5-brane \cite{Panda:2010uq} (based on \cite{Silverstein:2008sg,McAllister:2008hb,Flauger:2009ab}).
 The latter class of models are similar in spirit to the classical de Sitter vacua considered here, as only perturbative ingredients are introduced.

This paper is organized as follows. In Section \ref{sec:no-go-extrema},
we revisit the no-go theorems for de Sitter extrema.
Some form of these results were already obtained previously in the literature 
\cite{Hertzberg:2007wc,Haque:2008jz,Danielsson:2009ff,Wrase:2010ew}, 
but our approach is more systematic and suited for our subsequent discussions about the stability of these extrema. In Section \ref{sec:new-no-go}, we generalize the setups in Section \ref{sec:no-go-extrema} and
derive some new no-go theorems 
for (meta)stable de Sitter vacua.
We enumerate several ``minimal'' setups necessary to evade 
these no-goes.
We end with some discussions in Section \ref{sec:discussions}.
Some details are relegated to an appendix.

\section{No-go Theorems for de Sitter Extrema}\label{sec:no-go-extrema}

Here, we analyze the conditions for de Sitter extrema to arise in classical supergravity (with localized sources). 
Though some forms of these results were  previously obtained
\cite{Hertzberg:2007wc,Haque:2008jz,Danielsson:2009ff,Wrase:2010ew},
our result here is more complete, and our presentation will streamline our subsequent discussion in Section \ref{sec:new-no-go} on more complicated set-ups and on the stability of classical de Sitter vacua.

 Our analysis applies to both Type IIA and Type IIB string theories. Consider compactification to 4D, with the following ansatz for the metric:
 \begin{align}
  ds_{10}^2 =& \tau^{-2} ds_4^2 + \rho ds_6^2
  \label{metric ansatz}
 \end{align}
 where we took the Weyl factor to be $\tau = e^{-\phi} \rho^{3/2}$ such that the kinetic terms for the universal moduli $\rho$ and $\tau$ in the 4D Einstein frame do not mix.
 
 Various fluxes $H_3, F_p$, localized $q$-brane sources and the 6D curvature
 contribute to the 4D potential in some specific way:
 \begin{equation}
  V_{H_3} = A_{H_3} \tau^{-2} \rho^{-3}, \quad V_{F_p} = A_{F_p} \tau^{-4} \rho^{3-p}, \quad V_q = A_q \tau^{-3} \rho^{(q-6)/2}, \quad V_{R_6} = A_{R_6} \tau^{-2} \rho^{-1}.
 \end{equation}
 The coefficients $A_{H_3}$ and $A_{F_q}$ of the flux potentials are defined to be positive, while the coefficients $A_q$ of the $p$-branes contributions (including D-branes and O-planes) and $A_{R_6}$ for the curvature contribution  can be either positive or negative.
 Note that all these potentials go to zero when we take $\tau\rightarrow \infty$ while keeping the others finite.
 Therefore there always exist a Minkowski vacuum asymptotically.

Here we follow more closely the discussions in \cite{Hertzberg:2007wc,Wrase:2010ew} which analyze  the conditions for a de Sitter extremum to arise in this context.
 The idea is that if we can find an inequality of the following form:
 \begin{equation}
  \begin{split}
   &D \equiv - a \tau \partial_\tau - b \rho \partial_\rho,\\
   &D V \geq c V,
  \end{split}
  \label{old no-go for 1st derivative}
  \end{equation}
 with non-trivial real constants $a, b$, and $c>0$,
 then a positive energy extremum of the potential is excluded.
 To evade this no-go, one can enumerate a set of minimal ingredients needed.

 As a simple example, let's consider an effective 4D potential receiving contributions from $R_6, H_3$ and two more components taken from the set of $ F_{p}, {\rm O}q$.
 D-branes typically introduce additional open-string moduli, and thus as a first pass, we do not include them for simplicity.
 Here we analyze all possibilities which can evade the condition (\ref{old no-go for 1st derivative}), and then confirm that all candidates of this type were tabulated in \cite{Wrase:2010ew}.
 From now on we restrict ourselves to the following ingredients for simplicity:
 \begin{equation}
  \begin{split}
   &F_0, \ F_2, \ F_4, \ F_6,\ {\rm O}4, \ {\rm O}6 {\rm \ in \ IIA} \\
   &F_1, \ F_3,\  F_5,\  {\rm O}3, \ {\rm O}5, \ {\rm O}7 {\rm \ in \ IIB}
  \end{split}
  \label{potential ingredients}
 \end{equation}
 More complicated setups give us more examples, some excluded by
 the no-goes and some evading them.
 However since we will find examples of each type within this setup, we will concentrate on the above limited components.

 First we focus on the case with $R_6, \,H_3,\, F_p$, O$q$ (same as in \cite{Wrase:2010ew}).
 If the following conditions are satisfied:
 \begin{equation}
  \begin{split}
   &DV = (2a+b) V + 2b V_{H_3} + \left(2a + (p-4) b\right) V_{F_p} + \left( a + \left(2-{q \over 2} \right)b\right) V_q,\\
   &2a+b >0, \quad 2b\geq 0, \quad 2a + (p-4) b \geq 0, \quad a + \left(2-{q \over 2} \right)b\leq 0,
  \end{split}
 \end{equation}
we see that $DV \geq (2a+b) V$ and this leads to a no-go theorem for de Sitter extrema.
 However, the inequality can be violated if we have the following sources:
 \begin{equation}
  \begin{split}
   {\rm IIA}:&(F_0, {\rm O}4),\  (F_0, {\rm O}6),\ (F_2, {\rm O}4),\\
   {\rm IIB}:&(F_1, {\rm O}3),\ (F_1, {\rm O}5),\ (F_3, {\rm O}3),\ (F_5, {\rm O}3)
  \end{split}
  \label{old no no-go}
 \end{equation}
 with suitable sign of $R_6$ for each case, which we omit here but will be clarified later.
 A nontrivial $H_3$ is required in some cases, but we do not specify the details here.

 Upon a similar analysis but with $R_6, H_3, F_{p_1}, F_{p_2}$ and also $R_6, H_3, {\rm O}q_1, {\rm O}q_2$, we are left with two additional cases, i.e., $({\rm O}3, {\rm O}5), \ ({\rm O}3, {\rm O}7)$ which can potentially evade the no-go.
 However, a more detailed analysis showed that these two cases can evade the no-go only if the number of O5 (respectively O7) is zero.
 Therefore we have in total 7 cases listed in (\ref{old no no-go}), where the no-go does not apply.
 Note that evading the no-go here only means that the 4D potential can admit a de Sitter extremum, but does not guarantee the stability of such extremum.

 Now, if we include the contributions to the potential from the O8, O9-planes, the constraints for evading the no-goes leave us only with (O3, O9).
 However again this situation is possible only when the number of O9 is zero.
 Thus we can conclude that the result obtained in (\ref{old no no-go}) is most general, with two additional ingredients beyond $R_6$ and $H_3$.

 In the next section, we will proceed further to analyze the stability of the candidate de Sitter extrema.
 Although the ingredients presented in (\ref{old no no-go}) are minimal in terms of evading the no-go  (\ref{old no-go for 1st derivative}) for de Sitter extrema, all these minimal scenarios turn out to give only unstable extrema.
 Since the more complicated setups we consider in the next section subsume the simpler cases enumerated here, we will relegate our discussion of the no-go for stability to the next section.

\section{No-go Theorems for the Stability of de Sitter Extrema}\label{sec:new-no-go}
We now generalize the setup in the previous section to include three additional components
 beyond $R_6$ and $H_3$ and reanalyze the no-go theorems for de Sitter extrema 
 presented in  (\ref{old no-go for 1st derivative}).
  The three components are taken from the set:
  $F_p, {\rm O}q$,
  in particular, those listed in (\ref{potential ingredients}).
  Then evading the no-go for extremal (\ref{old no-go for 1st derivative}) leaves us with the following possibilities: 
  \begin{equation}
   \begin{split}
    {\rm IIA}:
    &(F_0, F_2, {\rm O}4), \ (F_0, F_4, {\rm O}4),\ (F_0, F_6, {\rm O}4), \ (F_2, F_4, {\rm O}4),\ (F_2, F_6, {\rm O}4), \\
    &(F_0, F_2, {\rm O}6),\ (F_0, F_4, {\rm O}6),\ (F_0, F_6, {\rm O}6), \ (F_0, {\rm O}4, {\rm O}6), \ (F_2, {\rm O}4, {\rm O}6),\\
    {\rm IIB}:
    &(F_1, F_3, {\rm O}3), \ (F_1, F_5, {\rm O}3), \ (F_3, F_5, {\rm O}3), \ (F_1, F_3, {\rm O}5), \ (F_1, F_5, {\rm O}5),\\
    &(F_1, {\rm O}3, {\rm O}5),\ (F_1, {\rm O}3, {\rm O}7), \ (F_3, {\rm O}3, {\rm O}5), \ (F_3, {\rm O}3, {\rm O}7), \ (F_5, {\rm O}3, {\rm O}5),\\
    &(F_5, {\rm O}3, {\rm O}7),\ (F_1, {\rm O}5, {\rm O}7),
   \end{split}
   \label{list for no no-go for extrema with three}
  \end{equation}
  We will analyze the stability of de Sitter extrema arising from all these cases.
  Again the sign of the 6D curvature should be chosen appropriately in each case, and to avoid cluttering our discussion, we omit such details here.

  Let us examine
   the (in)stability of the candidates listed in (\ref{list for no no-go for extrema with three}).
  We restrict our analysis to the  
   universal moduli subspace, as the positivity of the mass matrix 
   of this 2D subspace
   is a necessity condition for the absence of unstable modes in the full moduli space, according to the {\it Sylvester's criterion} in linear algebras
  \footnote{We are grateful to Thomas van Riet, Timm Wrase, and especially Marco Zagermann for bringing our attention  this point.}.
  The criterion can be stated as follows (see e.g. \cite{Gilbert:1991}):
  \begin{quotation}
   An $N \times N$ Hermitian (e.g., real-symmetric) matrix is positive-definite if and only if 
   the determinants of the upper-left $n \times n$ submatrices ($n \leq N$) are all positive, 
   or more mathematically precise, all of the leading principle minors are positive.
  \end{quotation}
  For instance,
  let us apply this criterion first to the $2 \times 2$ mass matrix $M$ of the $(\rho,\tau)$ subspace.
  The positivity of the determinant of $M$ requires the diagonal components to be both positive or both negative.
  In addition, the positivity of the upper-left most component is also required (by Sylvester's criterion) for $M$ to be positive definite, thus $\tr(M)>0$ and $\det (M) >0$.
  Now, applying Sylvester's criterion to the mass matrix of the full moduli space, we see that $\tr(M)>0$ together with $\det (M) >0$ are necessary conditions   for the full moduli mass matrix to be positive definite.

  The eigenvalues of the two-by-two mass matrix $M \equiv \partial_{\rho_i} \partial_{\rho_j} V|_{\rm ext}$ can be easily calculated:
  \begin{equation}
   \begin{split}
    {\rm eigenvalues}(M) =& {1\over 2} \left(\tr (M) \pm \sqrt{(\tr (M))^2 - 4 \det (M)}\right).
   \end{split}
   \label{conditions for stability}
  \end{equation}
   So a stable minimum can
    exist if both $\tr(M)>0$ and $ 0< \det (M) \leq (\tr (M))^2 /4$.
  We also consider the case in which we have a zero eigenvalue separately.
  In the following, we will show that in the minimal scenarios enumerated in (\ref{old no no-go}), there is at least one tachyonic or flat direction.
  Furthermore, in the remainder of this section, 
  we will show some no-go examples where $\tr(M)>0$ and $\det (M) >0$ cannot be simultaneously satisfied.
  Therefore, owing to the Sylvester's criterion, 
  the full moduli mass matrix must necessarily contain at least an unstable mode.

  We now analyze the stability of the setups enumerated in (\ref{list for no no-go for extrema with three}), which were already shown to evade the no-go for de Sitter extrema.
  Since  going through the details case by case is not very illuminating, 
  we will work out one case in detail here and relegate the details of all other cases to
  appendix \ref{sec:other-examples-no}.
  We will further enumerate in the next section
  the minimal ingredients for
  evading the {\it refined} no-go for de Sitter vacua we found here.

  Let's consider classical de Sitter solutions in Type IIA string theory 
  with $R_6, H_3, F_0, F_2$, and O4,
  as these ingredients were shown to evade the no-go for de Sitter extrema.
  As the critical point, one finds the following two constraints between the coefficients of the potential and the values of the moduli fields.
  \begin{equation}
   \begin{split}
    A_{H_3} =& - {A_{R_6} \rho^2 \over 7} + {13 A_{F_0} \rho^6 \over 7 \tau^2} + {A_{F_2} \rho^4 \over \tau^2 },\quad
    A_4 =- {4 A_{R_6} \tau \over 7} - {18 A_{F_0} \rho^4 \over 7 \tau} - {2 A_{F_2} \rho^2 \over \tau}.
   \end{split}
   \label{constraints for extremal}
  \end{equation}
  Usually we solve 
  for the moduli given the coefficients;
  here, however, we use the constraint equations
  to
  replace 
  the
  coefficients with moduli fields instead.
  This 
  does not mean that these coefficients are functions of the universal moduli, as these equalities  
  only hold on-shell.
  Upon substituting (\ref{constraints for extremal}), the potential at the extremum is:
  \begin{equation}
   \begin{split}
    V_{\rm ext} =& {2 A_{F_0} \rho^3 \over 7 \tau^4} + {2 A_{R_6} \over 7 \rho \tau^2}.
   \end{split}
  \end{equation}
  Here we see that
  a positive $A_{R_6}$, corresponding to a negatively curved manifold, helps to obtain a positive extremum of the potential, otherwise $A_{F_0}$ should be 
  sufficiently large to compensate for the contribution from a positive (or zero) internal curvature.

  We proceed to 
  analyze the stability of the extremum,
   following the argument around (\ref{conditions for stability}).
  Actually we do not need to 
  know the 
  trace of the mass matrix 
  in this example to show the no-go for 
  its
  stability.
  Instead, we focus on the determinant at the extrema which can be written as:
  \begin{equation}
   \begin{split}
   \det \left(\partial_{\rho_i} \partial_{\rho_j} V |_{\rm ext}\right)
    =& {4 \over 56 \rho^4 \tau^{10}} \left[\left(4A_{R_6} \tau^2 - 87 A_{F_0} \rho^4 - 35 A_{F_2} \rho^2\right)^2 \right. \\
    & \hspace{5em} \left.- 35 \left(35 A_{F_2}^2 \rho^4 + 182 A_{F_0} A_{F_2} \rho^6 + 243 A_{F_0}^2 \rho^8 \right)\right].
   \end{split}
  \end{equation}
  Since we are interested in
  $V_{\rm ext} >0,\  A_{H_3} \geq 0$, we are limited to the range:
   $- A_{F_0} \rho^4 < A_{R_6} \tau^2 \leq 7 A_{F_2} \rho^2 + 13 A_{F_0} \rho^4$.
  From the 
  explicit expression
  of the determinant,
  and the above inequalities, we found that
  the determinant is bounded from above by:
  \begin{equation}
   \det \left(\partial_{\rho_i} \partial_{\rho_j} V |_{\rm ext}\right)
    < - {16 A_{F_0}^2 \rho^4 \over \tau^{10}}.
  \end{equation}
  Therefore we see that the constraints, $V_{\rm ext}>0,\ \det \left(\partial_{\rho_i} \partial_{\rho_j} V |_{\rm ext}\right) \geq 0,\ A_{H_3} \geq0$ cannot be satisfied simultaneously, regardless of the sign of $R_6$.

  Let us analyze the case in which $\det \left(\partial_{\rho_i} \partial_{\rho_j} V |_{\rm ext}\right) =0$ while $\tr \left(\partial_{\rho_i} \partial_{\rho_j} V |_{\rm ext}\right) \geq0$, meaning that at least one eigenvalue of the mass matrix is zero.
  However we exclude the situations which can satisfy $\det \left(\partial_{\rho_i} \partial_{\rho_j} V |_{\rm ext}\right) > 0$, since then the examples have stable $dS$ vacua in some region. 
  Actually there is only one situation which have a zero eigenvalue, not positive, which is $(R_6, F_5, {\rm O3})$ without any other sources.
 In this case, the potential can be rewritten as
  \begin{equation}
   V(\sigma) = {A_{F_5} \over \sigma^4} + {A_3 \over \sigma^3} + {A_{R_6} \over \sigma^2},
    \label{potential with one flat direction}
  \end{equation}
  with the definition $\sigma = \tau \rho^{1/2}$.
  Since the non-trivial potential is generated only for the direction of $\sigma$, we have one remaining flat direction defined by $\delta = \tau \rho^{-1/2}$ in 2D moduli space.
  Therefore one modulus field $\delta$ cannot be stabilized, by the classical ingredients we consider.

  Applying similar analysis to all the cases in (\ref{list for no no-go for extrema with three}),
  we found that 
  Type II string theories in the presence of $R_6$, $H_3$, and only the following sources:
  \begin{equation}
   \begin{split}
    {\rm IIA}: & (F_0, F_2, {\rm O}4), \ (F_0, {\rm O}4, {\rm O}6), \\
    {\rm IIB}: & (F_1, F_3, {\rm O}3), \ (F_1, {\rm O}3, {\rm O}5), \ (F_1,  {\rm O}3, {\rm O}7), \ (F_3, {\rm O}3, {\rm O}7),\ (F_5, {\rm O}3, {\rm O}5), \ (F_5, {\rm O}3, {\rm O}7)
   \end{split}
   \label{list for new no-go}
  \end{equation}
  do not lead to metastable de Sitter vacua.
  For the cases in Type IIA listed above in (\ref{list for new no-go}), $V_{\rm ext} > 0$ and  $\det \left(\partial_{\rho_i} \partial_{\rho_j} V |_{\rm ext}\right) \geq 0$ cannot be satisfied simultaneously.
  While in Type IIB, we cannot satisfy $\tr \left(\partial_{\rho_i} \partial_{\rho_j} V |_{\rm ext}\right)\geq 0, \  \det \left(\partial_{\rho_i} \partial_{\rho_j} V |_{\rm ext}\right) \geq 0$ with $V_{\rm ext} >0$, except for the case $(R_6, F_5, {\rm O}3)$ in which we have one flat direction.
Thus, we found a new no-go theorem which states that
for the cases enumerated in (\ref{list for new no-go}),
when $V_{\rm ext} >0$, 
there is at least one tachyonic 
or flat direction.
  We also see that the ``minimal'' setups (enumerated in (\ref{old no no-go})) designed to evade the no-go for de Sitter extrema turn out to be all unstable.
  Even if we generalized these ``minimal'' setups in (\ref{old no no-go}) to include more contributions from fluxes and orientifold planes (see (\ref{list for no no-go for extrema with three})), some of these generalized setups still forbid stable positive minima.

 \section{Evading the no-goes\label{sec:no-no-goes}}
 Next, we
 consider the cases in (\ref{list for no no-go for extrema with three}) other than the ones in
 (\ref{list for new no-go}) which were already excluded in the previous section based on stability. 
 Among these many cases, we would like to classify what kinds of sources are essential in evading the no-go theorems for de Sitter vacua.
 Just like the previous section, we will work out one case in detail, and 
then simply provide a list of cases which can evade all the no-goes which we found after a thorough analysis. 
  
  Let's consider the setup with $R_6, H_3, F_0, F_2$, O6.
  At the extremum, the coefficients $A_{H_3}, A_6$, and the potential go like:
  \begin{equation}
   \begin{split}
    A_{H_3} =& - {A_{R_6} \rho^2 \over 3} + {A_{F_2} \rho^4 \over 3 \tau^2} + {A_{F_0} \rho^6 \over \tau^2}, \quad
    A_{6} = -{4 A_{R_6} \tau \over 9 \rho} - {14 A_{F_2} \rho \over 9 \tau} - {2 A_{F_0} \rho^3 \over \tau},\\
    V_{\rm ext} =& - {2 A_{F_2} \rho \over 9 \tau^4} + {2 A_{R_6} \over 9 \rho \tau^2}.
   \end{split}
  \end{equation}
  Then we can analyze the 
  stability of the de Sitter critical point from
  the  trace and the determinant  of the mass matrix at the extrema:
  \begin{equation}
   \begin{split}
    \tr \left(\partial_{\rho_i} \partial_{\rho_j} V |_{\rm ext}\right)
    =&{2 \over 3\rho^3 \tau^6} \left(5 A_{F_2} \rho^4 + 3 A_{F_0} \rho^6 + 6 A_{F_2} \rho^2 \tau^2 - 2 A_{R_6} \rho^2 \tau^2 + 27 A_{F_0} \rho^4 \tau^2 - 3 A_{R_6} \tau^4 \right),\\
    \det \left(\partial_{\rho_i} \partial_{\rho_j} V |_{\rm ext}\right)
    =&{4 \over 3 \rho^4 \tau^{10}} \left(7 A_{F_2}^2 \rho^4 + 33 A_{F_0} A_{F_2} \rho^6 - 9 A_{F_2} A_{R_6} \rho^2 \tau^2 - 21 A_{F_0} A_{R_6} \rho^4 \tau^2 + 2 A_{R_6}^2 \tau^4 \right).
   \end{split}
  \end{equation}
  To obtain the validity range for de Sitter minima, we need to satisfy $\tr \left(\partial_{\rho_i} \partial_{\rho_j} V |_{\rm ext}\right) >0, \ 0< \det \left(\partial_{\rho_i} \partial_{\rho_j} V |_{\rm ext}\right) \leq (\tr \left(\partial_{\rho_i} \partial_{\rho_j} V |_{\rm ext}\right))^2/4$, simultaneously with $V_{\rm ext} >0,\  A_{H_3} \geq 0,\  A_{6} < 0$.
  After some 
  simplifications, these inequalities leave us with the following range of parameters:
  \begin{equation}
   \begin{split}
    {A_{F_2} \rho^2 \over \tau^2} < A_{R_6} < {11 A_{F_2} \rho^2 \over 7 \tau^2}, \quad
    A_{F_0} > { -7 A_{F_2}^2 \rho^4 + 9 A_{F_2} A_{R_6} \rho^2 \tau^2 - 2 A_{R_6}^2 \tau^4 \over 33 A_{F_2}\rho^6 - 21 A_{R_6} \rho^4 \tau^2},
   \end{split}
   \label{range for stable dS 1}
  \end{equation}
  which can be satisfied, and thus the no-go theorem for the stability of de Sitter extrema can be evaded.
  Note that $A_{H_3} >0$ for the parameter region above.

  Since we are interested in finding the parameter regime which gives stable de Sitter vacua,
    we omitted the case when $\det \left(\partial_{\rho_i} \partial_{\rho_j} V |_{\rm ext}\right) =0$
  which can be satisfied 
    non-trivially in this example if  the second inequality in (\ref{range for stable dS 1}) becomes an equality.
  If an eigenvalue of the mass matrix is zero, we need to check higher order terms for stability.

  One can repeat this analysis for all the other cases.
  Since the details are not very illuminating, we simply state our results as follows:

 \noindent $(i)$  In Type IIA string theory, we can satisfy
 the conditions for the existence of de Sitter critical points, and the requirements of stability in the following setups:
  \begin{itemize}
   \item $R_6, H_3, F_0, F_2$, O6 with $A_{R_6} >0,\ A_{H_3} >0$,
   \item $R_6, H_3, F_0, F_4$, O6 with $A_{R_6} >0, \ A_{H_3}\geq 0$,
   \item $R_6, H_3, F_0, F_6$, O6 with $A_{R_6} >0,\ A_{H_3}\geq 0$,
   \item $R_6, H_3, F_0, F_4$, O4 with $A_{R_6} \gtrless 0,\ A_{H_3} >0$ or $A_{R_6} >0,\ A_{H_3} =0$,
   \item $R_6, H_3, F_0, F_6$, O4 with $A_{R_6} \gtrless 0,\ A_{H_3} >0$ or $A_{R_6} >0,\ A_{H_3} =0$,
   \item $R_6, H_3, F_2, F_4$, O4 with $A_{R_6} > 0,\ A_{H_3}\geq 0$,
   \item $R_6, H_3, F_2, F_6$, O4 with $A_{R_6} >0,\ A_{H_3} \geq 0$,
   \item $R_6, H_3, F_2$, O4, O6 with $A_{R_6} >0,\ A_{H_3} >0$.
  \end{itemize}  
\noindent $(ii)$  In Type IIB string theory,  the following setups can admit stable de Sitter minimum:
  \begin{itemize}
   \item $R_6, H_3, F_3, F_5$, O3 with $A_{R_6} >0,\ A_{H_3}>0$,
   \item $R_6, H_3, F_1, F_5$, O3 with $A_{R_6} \gtrless 0,\ A_{H_3}>0$,
   \item $R_6, H_3, F_1, F_5$, O5 with $A_{R_6} >0,\ A_{H_3} \geq 0$,
   \item $R_6, H_3, F_1, F_3$, O5 with $A_{R_6} > 0,\ A_{H_3}>0$,
   \item $R_6, H_3, F_3$, O3, O5 with $A_{R_6} >0,\ A_{H_3}>0$,
   \item $R_6, H_3, F_1$, O5, O7 with $A_{R_6} >0, \ A_{H_3} >0$.
  \end{itemize}
  As we increase the number of ingredients (fluxes and localized sources), we are likely to find more and more examples that evade the no-goes (for both the existence of an  extremum and its stability), but the above list constitutes the ``minimal'' setups for classical de Sitter vacua.
  Interestingly, the no-goes can be evaded with compactifications on a
  {\it positively curved} or a {\it flat} manifold in the presence of $H_3$ flux, even though in most cases, a negatively curved manifold is preferred.

  \section{Discussions}\label{sec:discussions}

  In this paper, we 
  have presented several new no-go theorems for classical de Sitter vacua, i.e., de Sitter constructions using only 6D curvature, fluxes and O-planes in Type IIA and IIB string theories.
  In addition to the no-goes for de Sitter extrema 
  previously derived in this context \cite{Hertzberg:2007wc,Haque:2008jz,Danielsson:2009ff,Wrase:2010ew},
  we found that
  constraints on the stability of these extrema can 
  further
  eliminate a significant portion of the landscape.
  We enumerate the {\it minimal} ingredients
  needed to evade these no-goes.
  Most of these minimal setups we found involve negatively curved 6D manifolds as originally suggested in \cite{Silverstein:2007ac}, though there still remain several interesting possibilities with positively curved 6D manifolds.
  It would be interesting to see if such minimal setups can indeed be realized in terms of explicit models.

  Recently it was argued that warping and/or stringy corrections are necessarily important for compactifications on manifolds whose curvature is everywhere negative \cite{Douglas:2010rt}.
  This is because the equations of motion cannot be satisfied pointwise in an
 everywhere negatively curved internal space if the only negative tension objects at our disposal, i.e. the orientifold planes are localized. 
 The universal K\"ahler moduli defined in (\ref{metric ansatz}) is that of an unwarped case, but is modified in the presence of warping \cite{Giddings:2005ff,Kodama:2005cz,Kodama:2006ay,Koerber:2007xk,Shiu:2008ry,Douglas:2008jx,Martucci:2009sf,Chen:2009zi}
(see, in particular \cite{Frey:2008xw} which is more suited for our present discussion).
 In this work, we sidestep these issues of warping by implicitly smearing the orientifold planes, which can be thought of as an approximation before the fully backreacted solutions are found
 \footnote{We consider smearing just on a base manifold transverse to the O$q$-planes, such that the O$q$-planes do not acquire additional moduli dependences as compared to a localized one.}.
  Furthermore,
 since our analysis is carried out in the 4D effective field theory where the internal space is integrated out.
 Therefore, it would also apply to manifolds which are not everywhere negatively curved, but with an overall negative 6D curvature.
We should emphasize that while we found the necessary constraints for stable classical $dS$ minima, there is no guarantee that explicit backgrounds satisfying the requirements exist. Moreover, the full backreaction of several localized sources remains an open challenging issue.

We focus our search for de Sitter vacua whose constructions do not invoke explicit SUSY breaking localized sources. Introducing sources such as anti-branes would certainly enlarge the list of possibilities. 
For example, the no-go theorem for the stability of de Sitter extremum constructed from IIB string theory with $R_6$, $H_3$, $F_0$, O4, and O6 can be evaded by replacing O4 with D4-$\overline{\rm D4}$ pairs.
In fact, the no-go theorems can similarly be evaded for $R_6$, $H_3$, $F_1$, D3-$\overline{\rm D3}$, O5, also for $R_6$, $H_3$, $F_5$, O3, D5-$\overline{\rm D5}$ and $R_6$, $H_3$, $F_5$, O3, D7-$\overline{\rm D7}$.
 Pairs of localized D$3$-$\overline{{\rm D}3}$ \cite{Kachru:2002gs}
 were also used for uplifting the $AdS$ vacua to $dS$ in \cite{Kachru:2003aw}.
 Recently the backreaction of such D$3$-$\overline{{\rm D}3}$ pairs in the Klebanov-Strassler throat were discussed in \cite{DeWolfe:2008zy,McGuirk:2009xx,Bena:2009xk} (see also further discussions in \cite{Bena:2010ze,Dymarsky:2011pm,Bena:2011wh}).
 Related studies on the backreaction of localized sources in simpler setups (though more closely related to the classical de Sitter vacua discussed here) were considered in \cite{Blaback:2010sj,Blaback:2011nz}.

  \section*{Acknowledgment}
  We have benefited from discussions with Ulf Danielsson, Hirotaka Hayashi, Masazumi Honda, Shajid Haque, Paul Koerber, Thomas Van Riet, S.-H. Henry Tye, Sandip P. Trivedi, Timm Wrase and Marco Zagermann.
  GS and YS thank the Hong Kong Institute for Advanced Study greatly for their hospitality.
GS is supported in part by a DOE grant under contract DE-FG-02-95ER40896, and a Cottrell Scholar Award from Research Corporation.

\appendix
 \section{Instability Analysis for Other Examples  \label{sec:other-examples-no}}
 This appendix is devoted to complete the 
 stability analysis for all the cases in (\ref{list for new no-go}),
 which were skipped in 
 section \ref{sec:new-no-go}, and demonstrate that they all lead to a no-go for stable de Sitter vacua.

  \subsection{$R_6, H_3, F_0,$ O4, O6 in IIA}
 The requirements of an extremum lead to the following two conditions:
 \begin{equation}
   A_{H_3} = - {A_{R_6} \rho^6 \over 3} + {A_{F_0} \rho^6 \over \tau^2} - {A_4 \rho^2 \over 3 \tau}, \quad
    A_{6} = - {2A_{F_0} \rho^3 \over \tau} - {4 A_{R_6} \tau \over 9 \rho} - {7 A_4 \over 9 \rho}.
 \end{equation}
  Then the potential 
  at its extremum, upon substituting $A_{H_3}$, and $A_6$, becomes
  \begin{equation}
   V_{\rm ext} = {2 A_{R_6} \over 9 \rho \tau^2} - {A_4 \over 9 \rho \tau^3}.
  \end{equation}
  Therefore we see that a negatively curved manifold together with a number of O4-planes,
  or with a certain amount of $F_0$ flux
  help to obtain $dS$ extrema.

  The determinant of the mass matrix at the extremum becomes
  \begin{equation}
   \begin{split}
    \det \left(\partial_{\rho_i} \partial_{\rho_j} V |_{\rm ext}\right)
    = {1 \over 3 \rho^4 \tau^9}  \left(60 A_4 A_{F_0} \rho^4 - 7 A_4^2 \tau - 84 A_{F_0} A_{R_6} \rho^4 \tau + 4 A_4 A_{R_6} \tau^2 + 8 A_{R_6}^2 \tau^3 \right) .
   \end{split}
  \end{equation}
 Thus, we cannot satisfy $V_{\rm ext} > 0, \ \det \left(\partial_{\rho_i} \partial_{\rho_j} V |_{\rm ext}\right) \geq 0, \ A_{H_3} \geq 0, \ A_4 \leq 0$ simultaneously.
  This means there is at least one tachyonic direction at the positive extremum.

  \subsection{$R_6, H_3, F_1, F_3$, O3 in IIB}
The conditions on extrema lead to the following potential at the critical point: 
  \begin{equation}
   \begin{split}
    A_{H_3} =& {A_{F_3} \rho^3 \over \tau^2} + {2 A_{F_1} \rho^5 \over \tau^2}, \quad
    A_3 = -{2 A_{F_3} \rho^{3/2} \over \tau} - {8 A_{F_1} \rho^{7/2} \over 3 \tau} - {2\over 3} {A_{R_6} \rho^{1/2} \tau},\\
    V_{\rm ext} =& {A_{F_1} \rho^2 \over 3 \tau^4} + {A_{R_6} \over 3 \rho \tau^2}.
   \end{split}
  \end{equation}
  The trace and the determinant of the mass matrix go like
  \begin{equation}
   \begin{split}
   \tr \left(\partial_{\rho_i} \partial_{\rho_j} V |_{\rm ext}\right)
    =& {1\over 2\rho^3 \tau^6}  \left(4 A_{F_3} \rho^3 + 9 A_{F_3} \rho \tau^2 - 4 A_{R_6} \rho^2 \tau^2 - A_{R_6} \tau^4 + 32 A_{F_1} \rho^3 \tau^2 \right),\\
   \det \left(\partial_{\rho_i} \partial_{\rho_j} V |_{\rm ext}\right)
    =& - {16 \over \rho^3 \tau^{10}} \left(4 A_{F_1}^2 \rho^5 + A_{F_3} A_{R_6} \tau^2 + A_{F_1} \rho^2 (A_{F_3} \rho + 3 A_{R_6} \tau^2) \right),
   \end{split}
  \end{equation}
   We see that the conditions $V_{\rm ext}>0,\ \det \left(\partial_{\rho_i} \partial_{\rho_j} V |_{\rm ext}\right) \geq 0, \  \tr \left(\partial_{\rho_i} \partial_{\rho_j} V |_{\rm ext}\right) \geq 0,\  A_{H_3} \geq0$ cannot be satisfied simultaneously.

  \subsection{$R_6, H_3, F_1$, O3, O5 in IIB}
At the extremum of the potential:
  \begin{equation}
   \begin{split}
    A_{H_3} =& {2 A_{F_1} \rho^5 \over \tau^2} + {A_5 \rho^{5/2} \over 2 \tau}, \quad
    A_3 = - {8 A_{F_1} \rho^{7/2} \over 3 \tau} - {2 A_{R_6} \rho^{1/2} \tau \over 3} - {4 A_5 \rho \over 3},\\
    V_{\rm ext} =& {A_{F_1} \rho^2 \over 3 \tau^4} + {A_{R_6} \over 3 \rho \tau^2} + {A_5 \over 6 \rho^{1/2} \tau^3},
   \end{split}
  \end{equation}
  while the trace and the determinant of the mass matrix are found to be
  \begin{equation}
   \begin{split}
    \tr \left(\partial_{\rho_i} \partial_{\rho_j} V |_{\rm ext}\right)
    =& {1\over 4 \rho^3 \tau^5} \left(-4 A_5 \rho^{5/2} + 7 A_5 \rho^{1/2} \tau^2 - 8 A_{R_6} \rho^2 \tau -2 A_{R_6} \tau^3 + 64 A_{F_1} \rho^3 \tau \right),\\
    \det \left(\partial_{\rho_i} \partial_{\rho_j} V |_{\rm ext}\right)
    =& - {2 \over \rho^{7/2} \tau^{10}} (8 A_{F_1} \rho^{5/2} + A_5 \tau) \left(4 A_{F_1} \rho^3 + 2 A_5 \rho^{1/2} \tau + 3 A_{R_6} \tau^2 \right).
   \end{split}
  \end{equation}
  Again we cannot satisfy $V_{\rm ext}>0,\ \det \left(\partial_{\rho_i} \partial_{\rho_j} V |_{\rm ext}\right) \geq 0, \  \tr \left(\partial_{\rho_i} \partial_{\rho_j} V |_{\rm ext}\right) \geq 0,\  A_{H_3} \geq0,\  A_3\leq 0$ simultaneously regardless of the sign of $R_6$.

  \subsection{$R_6, H_3, F_5$, O3, O5 in IIB}
At the extremum of the potential:
  \begin{equation}
   \begin{split}
    A_{H_3} =& {A_5 \rho^{5/2} \over 2 \tau}, \quad
    A_3 = - {4 A_{F_5} \over 3 \rho^{1/2} \tau} - {2 A_{R_6} \rho^{1/2} \tau \over 3} - {4 A_5 \rho \over 3},\\
    V_{\rm ext} =& - {A_{F_5} \over 3 \rho^2 \tau^4} + {A_{R_6} \over 3 \rho \tau^2} + {A_5 \over 6 \rho^{1/2} \tau^3}.
   \end{split}
  \end{equation}
  then the trace and the determinant of the mass matrix becomes
   \begin{equation}
    \begin{split}
     \tr \left(\partial_{\rho_i} \partial_{\rho_j} V |_{\rm ext}\right)
     =&{1\over 4 \rho^4 \tau^6} \left(16 A_{F_5} \rho^2 + 4 A_{F_5} \tau^2 - 4 A_5 \rho^{7/2} \tau + 7 A_5 \rho^{3/2} \tau^3 - 8 A_{R_6} \rho^3 \tau^2 - 2 A_{R_6} \rho \tau^4 \right),\\
   \det \left(\partial_{\rho_i} \partial_{\rho_j} V |_{\rm ext}\right)
    =& - {2 A_5 \left(-6 A_{F_5} +3 A_{R_6} \rho \tau^2 + 2 A_5 \rho^{3/2} \tau\right) \over \rho^{9/2} \tau^9}.
    \end{split}
  \end{equation}
  In this case, we cannot satisfy $V_{\rm ext}>0,\ \det \left(\partial_{\rho_i} \partial_{\rho_j} V |_{\rm ext}\right) > 0, \  \tr \left(\partial_{\rho_i} \partial_{\rho_j} V |_{\rm ext}\right) \geq 0,\  A_{H_3} \geq0, \ A_5 \leq0$ simultaneously.
  When we satisfy $\det \left(\partial_{\rho_i} \partial_{\rho_j} V |_{\rm ext}\right)=0, \ \tr \left(\partial_{\rho_i} \partial_{\rho_j} V |_{\rm ext}\right) \geq 0$, the system is reduced to  $(R_6, F_5, {\rm O}3)$ which was showed to have one flat direction in (\ref{potential with one flat direction}).

  \subsection{$R_6, H_3,  F_1,$ O3, O7 in IIB}
  At the extremum of the 4D potential:
  \begin{equation}
   \begin{split}
    A_{H_3} =& {2 A_{F_1} \rho^5 \over \tau^2} + {A_7 \rho^{7/2} \over \tau}, \quad
    A_3 = -{8 A_{F_1} \rho^{7/2} \over 3 \tau} -{2 A_{R_6} \rho^{1/2} \tau \over 3}- {5 A_7 \rho^2 \over 3},\\
    V_{\rm ext} =& {A_{F_1} \rho^2 \over 3 \tau^4} + {A_7 \rho^{1/2} \over 3 \tau^3} + {A_{R_6} \over 3 \rho \tau^2}.
   \end{split}
  \end{equation}
  Then the trace and the determinant of  the mass matrix go like
  \begin{equation}
   \begin{split}
    \tr \left(\partial_{\rho_i} \partial_{\rho_j} V |_{\rm ext}\right)
    =& {1\over 2\rho^3 \tau^5} \left(32 A_{F_1} \rho^3 \tau - 4 A_{R_6} \rho^2 \tau - A_{R_6} \tau^3 -4 A_7 \rho^{7/2} + 11 A_7 \rho^{3/2} \tau^2 \right),\\
   \det \left(\partial_{\rho_i} \partial_{\rho_j} V |_{\rm ext}\right)
    =& {-4 \over \rho^{5/2} \tau^{10}}
    \left(16 A_{F_1}^2 \rho^{9/2} + A_7 \tau^2 (5 A_7 \rho^{3/2} + 4 A_{R_6} \tau) + 4 A_{F_1} (5 A_7 \rho^3 \tau + 3 A_{R_6} \rho^{3/2} \tau^2)\right).
   \end{split}
  \end{equation}
 It can be shown that $V_{\rm ext}>0,\ \det \left(\partial_{\rho_i} \partial_{\rho_j} V |_{\rm ext}\right) \geq 0, \  \tr \left(\partial_{\rho_i} \partial_{\rho_j} V |_{\rm ext}\right) \geq 0,\  A_{H_3} \geq0$ cannot be satisfied.

  \subsection{$R_6, H_3, F_3,$ O3, O7 in IIB}
 At the extremum of the 4D potential:
  \begin{equation}
   \begin{split}
    A_{H_3}=& {A_{F_3} \rho^3 \over \tau^2} + {A_7 \rho^{7/2}\over \tau}, \quad
    A_3 = - {2 A_{F_3} \rho^{3/2} \over \tau} - {2 A_{R_6} \rho^{1/2} \tau \over 3} - {5 A_7 \rho^2 \over 3},\\
    V_{\rm ext} =& {A_7 \rho^{1/2} \over 3 \tau^2} + {A_{R_6} \over 3 \rho \tau^2}
   \end{split}
  \end{equation}
  Then the determinant of the mass matrix becomes
  \begin{equation}
   \begin{split}
    \tr \left(\partial_{\rho_i} \partial_{\rho_j} V |_{\rm ext}\right)
    =& {1\over 2\rho^3 \tau^6} \left( 4 A_{F_3} \rho^3 + 9 A_{F_3} \rho \tau^2- 4 A_7 \rho^{7/2} \tau + 11 A_7 \rho^{3/2} \tau^3  - 4 A_{R_6} \rho^2 \tau^2 - A_{R_6} \tau^4\right),\\
   \det \left(\partial_{\rho_i} \partial_{\rho_j} V |_{\rm ext}\right)
    =& -{4\over \rho^3 \tau^9} \left(4 A_{F_3} A_{R_6} \tau + 5 A_7^2 \rho^2 \tau + 4 A_7 (A_{F_3} \rho^{3/2} + A_{R_6} \rho^{1/2} \tau^2)\right).
   \end{split}
  \end{equation}
  Thus we cannot satisfy $V_{\rm ext}>0,\ \det \left(\partial_{\rho_i} \partial_{\rho_j} V |_{\rm ext}\right) \geq 0, \  \tr \left(\partial_{\rho_i} \partial_{\rho_j} V |_{\rm ext}\right) \geq 0,\  A_{H_3} \geq0$ simultaneously.

  \subsection{$R_6, H_3, F_5,$ O3, O7 in IIB}
  At the extremum of the 4D potential:
  \begin{equation}
   \begin{split}
    A_{H_3} =& {A_7 \rho^{7/2} \over \tau}, \quad
    A_3 = - {4 A_{F_5} \over 3 \rho^{1/2} \tau} - {2 A_{R_6} \rho^{1/2} \tau \over 3} - {5 A_7 \rho^2 \over 3},\\
    V_{\rm ext} =& - {A_{F_5} \over 3 \rho^2 \tau^4} + {A_{R_6} \over 3 \rho \tau^2} + {A_7 \rho^{1/2} \over 3 \tau^3}.
   \end{split}
  \end{equation}
  We see immediately that the only allowed situation is $A_{H_3} = A_7 =0$ for O7, a case subsumed in the analysis around (\ref{potential with one flat direction}) which has at least one flat direction.

\providecommand{\href}[2]{#2}\begingroup\raggedright\endgroup


\end{document}